# Raman-Kerr Comb Generation Based on Parametric Wave Mixing in Strongly Driven Raman Molecular Gas Medium

Aurélien Benoît,[1] Anton Husakou,[2] Benoît Beaudou,[3] Benoît Debord,[1,3] Frédéric Gérôme,[1,3] and Fetah Benabid[1,3,*]

[1]*GPPMM Group, Xlim Research Institute, CNRS UMR 7252, University of Limoges, Limoges, France*
[2]*College of Max Born Institute, Forschungsverbund Berlin e.V.RudowerChaussee 1712489 Berlin, Germany*
[3]*GLOphotonics SAS, 123 avenue Albert Thomas 87060 Limoges Cedex, France*
[*]*Corresponding author: f.benabid@xlim.fr*

We report on experimental and theoretical demonstration of an optical comb spectrum based on a combination of cascaded stimulated Raman scattering and four-wave mixing mediated by Raman-induced nonresonant Kerr-type nonlinearity. This combination enabled to transform a conventional quasi-periodic Raman comb into a comb with a single and smaller frequency spacing. This new phenomenon is realized using a hollow-core photonic crystal fiber filled with 40 bars of deuterium, and pumped with a high-power picosecond laser. The resultant comb shows more than 100 spectral lines spanning over 220 THz from 800 nm to 1710 nm, with a total output power of 7.1 W. In contrast to a pure Raman comb, a 120 THz wide portion of the spectrum exhibits denser and equally-spaced spectral lines with a frequency spacing of around 1.75 THz, which is much smaller than the lowest frequency of the three excited deuterium Raman resonances. A numerical solution of the generalized nonlinear Schrödinger equation in the slowly varying envelope approximation provides very good agreement with the experimental data. The additional sidebands are explained by cascaded four-wave mixing between pre-existing spectral lines, mediated by the large Raman induced optical nonlinearity. The results open a new route to the generation of optical frequency combs that combine large bandwidth and high power controllable frequency spacing.

## I. INTRODUCTION

The generation of optical combs is turning into an extremely important topic in several scientific and technological fields, such as waveform synthesis [1], the optical clock [2], communication systems [3] or biomolecular imaging [4]. Historically, generation of optical combs was associated with mode-locked lasers [5] or with high harmonic generation using high-peak-power and ultra-short pulse laser sources [6-7]. The former has matured to be the cornerstone in optical frequency metrology thanks to the outstanding frequency stability of the comb spectral components [8], and the latter is a powerhouse in synthesizing attosecond pulse in VUV and Soft X-ray thanks to its multi-octave bandwidth [9]. During the last two decades, photonics' solutions to generate optical combs have emerged with potential to address current limitations such as compactness or multi-octave comb generation in more convenient spectral ranges. Among these, we list Kerr optical nonlinearity based combs, which are generated by micro-resonators, and pumped with a continuous-wave laser [10]. Here, the generated spectrum can be up to one octave wide and the frequency spacing of its components, set by the micro-resonator diameter, ranges from a few hundred GHz up to 1 THz. Another means of generating discrete spectrum is based on stimulated Raman scattering (SRS) in gas-filled hollow-core photonics-crystal-fiber (HC-PCF) [11]. Compared to mode-locked laser or micro-resonator mechanisms, the Raman-gas-filled HC-PCF based combs stand out with their large bandwidth [11], which can span more than five octaves from the ultraviolet to mid-infrared range, and with their high optical power and energy handling [12]. The generated spectra using Raman-gas-filled HC-PCF with nanosecond or picosecond pump pulses show intra-pulse phase coherence [13] and are typically composed of lines with several frequency spacings exclusively set by the rotational and/or vibrational Raman resonances [11-12, 14-15]. Consequently, the typical spectral structure of a HC-PCF based Raman comb consists of a central frequency $v_0$ set by the pump laser, and a cascade of sidebands with frequencies in the form of $v_{l,m,n} = v_0 + lv_{R1} + mv_{R2} + nv_{R3}$ for the case of a medium with three Raman transitions. Here, $l, m, n$ are integers corresponding to the negative (Stokes) and positive (anti-Stokes) orders of the Raman transitions with frequencies $v_{R1}$, $v_{R2}$ and $v_{R3}$ respectively. Note that the spectral structure of such SRS comb is set by resonant third-order susceptibility in the vicinity of the Raman resonance [16], with no measurable influence from the non-resonant non-linear susceptibility (*i.e.* Kerr effect) [17]. As such, the HC-PCF based Raman comb suffers from two major drawbacks that hinder their use in a number of applications. It exhibits multiple frequency spacing (quasi-periodic comb), and its Raman frequency is too large (> 10 THz for most of the representative molecular gases).

In this paper, we show a new approach to overcome this limit of Raman combs and to generate a small and single frequency-spacing comb in Raman gas media of the form $v_q = v_0 + q\Delta$. The approach relies on using a Raman gas containing at least two nearly commensurable resonance frequencies, so that, for example, $v_{R2}$ and $v_{R3}$ can be written as $v_{R2} = N\Delta + \delta$ and $v_{R3} = M\Delta + \delta'$ with $N$ and $M$ being positive integers, $\delta,\delta' \ll \Delta$. The proof of principle of the approach is undertaken by experimentally and theoretically demonstrating the generation of a wide comb in a Raman gas-filled HC-PCF that exhibits spectral lines with a frequency-spacing as low as 1.75 THz. The observed spectrum, which is generated from a high-pressure deuterium molecular gas, consists of up to 100 spectral lines covering a width of 220 THz. In addition to lines associated with the three excited Raman resonances of deuterium at frequencies of 89.6 THz, 12.4 THz and 5.3 THz respectively, a good part of spectrum includes a denser comb of 52 lines covering a range of 120 THz with a frequency spacing of 1.75 THz. We attribute these denser spectral lines to an action of Raman-induced parametric four wave-mixing (FWM) (*i.e.* the out-of-resonance contribution of the Raman susceptibility), and to a FWM process between close Raman sidebands to generate frequency components outside the Raman resonance frequencies.

## II. EXPERIMENTAL SETUP AND RESULTS

The experimental set-up consists of a high-power picosecond pulsed Yb-fiber laser coupled into a photonic microcell (PMC) comprising a length of HC-PCF whose ends are attached to gas cells [18]. The laser emits at 1030 nm, has an average power of 16 W, a repetition rate of 1 MHz, pulse duration of 27 ps and a self-phase modulation broadened spectrum with a bandwidth of 200 GHz [19]. The PMC is filled with deuterium molecules ($D_2$) at a pressure of 40 bar. The choice of $D_2$ is motivated by the frequency values of its two ro-vibrational Raman transitions. At room temperature, $D_2$ gas is a 2:1 molecular mixture of ortho-deuterium (o-$D_2$) and para-deuterium (p-$D_2$). It exhibits three Raman resonances: a vibrational one around $v_{R1}$ = 89.6 THz with a steady-state Raman gain of 0.37 cm/GW, and two ro-vibrational ones at $v_{R2}$ = 12.44 THz from o-$D_2$ [20], and at $v_{R3}$ = 5.33 THz from p-$D_2$ [21] with gains of ~0.12 cm/GW and ~0.05 cm/GW respectively. As a result, the deuterium Raman shifts for para- and ortho- molecules are close to be commensurable, with their ratio being equal to 3:7 within less than 0.1% of relative error. As such, we have 5.33 THz ~ 3Δ, and 12.44 THz ~ 7Δ, where Δ = ~ 1.78 THz coincides with the $|v_{R2} - 2v_{R3}|$ value, which is also the frequency spacing between the Raman lines $v_{l,m+1,-1}$ and $v_{l,m,1}$, and is three times smaller than the smallest Raman frequency. The impact of this peculiarity will be apparent later. Furthermore, the gains of the o-$D_2$ and p-$D_2$ ro-vibrational resonances have closer values (2:1 ortho- to para- ratio of gains) than for the hydrogen gas (3:1 ratio) [22], favoring thus the excitation of p-$D_2$ Raman resonance when compared to $H_2$. Finally, operating at a pressure as high as 40 bar ensures higher Raman gain but broader gain linewidth, typically ~ 10 GHz. The latter is notably useful to fulfilling the above-mentioned commensurability. The PMC fiber consists of a 3 m long inhibited-coupling single-ring tubular-lattice HC-PCF [23] with an inner diameter of 40 μm [Fig. 1(a)].

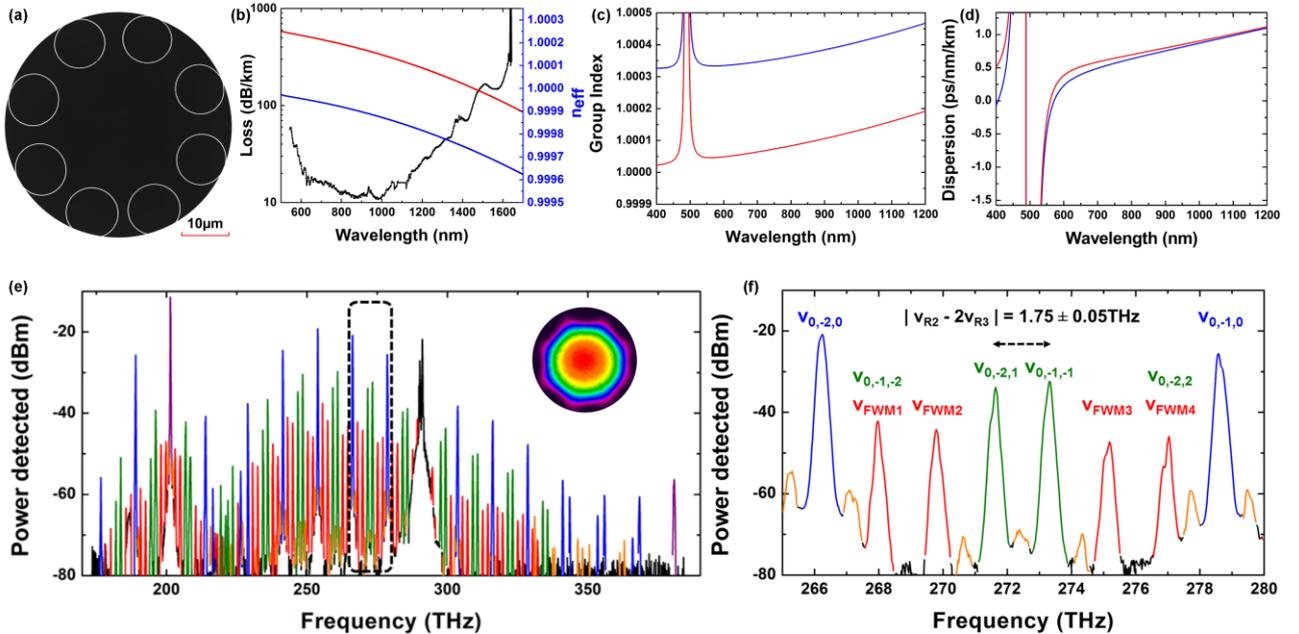

**Fig. 1.** (a) Scanning electron microscope of the inhibited-coupling single-ring tubular-lattice HC-PCF used for this experiment. (b) Evolution of losses (black curve) and effective refractive index in the vacuum (blue curve) and hydrogen (red curve) as a function of wavelength. (c) Evolution of group index as a function of the wavelength in the vacuum (blue curve) and in hydrogen (red curve). (d) Evolution of dispersion as a function of wavelength in the vacuum (blue curve) and in hydrogen (red curve). (e) Output spectrum after a 16 W power pump propagation in 3 m long HC-PCF filled with 40 bars of deuterium. The pump laser line is identified by black color, the vibrational Stokes and anti-Stokes by purple, the Stokes and anti-Stokes lines due to ortho-ro-vibrational Raman by blue, the Stokes and anti-Stokes lines due to para-ro-vibrational Raman by green, and four-wave-mixing generated sidebands by red. Inset: near field intensity distribution of the output beam. (f) Zoomed-in spectrum of the black-dashed rectangle.

The fiber transmission loss in the visible-infrared range is characterized by an ultrawide (over one octave) band from 600 to 1200 nm with loss around 11 dB/km [Fig. 1(b)]. Figure. 1(b) presents also the evolution of the refractive index as a function of the wavelength. Figs. 1(c-d) show the used fiber spectra of the group index and the dispersion respectively. A detailed account of all the optical properties of the HC-PCF is given in [23]. The laser beam is coupled into the PMC in a manner to match its beam size with the fiber fundamental core mode. Inspection of the fiber output beam profile over the operating spectral range of 800-1600 nm shows single mode propagation [inset of Fig. 1(e)]. Figure 1(e) presents the typical PMC output spectrum when excited with 16 W average power pump laser. The power of the generated and transmitted spectrum was measured to be 7.1 W. The spectrum spans a wide range from 170 to 390 THz and exhibits a large number of lines in a comb-like structure with more than 100 lines. The spectral components reveal that two purple-colored lines correspond in their frequency shifts (89.63 THz) to the vibrational Stokes $v_{-1,0,0}$ and anti-Stokes $v_{1,0,0}$ lines of the pump. The vibrational comb lines are accompanied by the comb of blue lines, which appear around both pump and the 1st order vibrational Stokes and anti-Stokes lines. The blue lines are readily associated with ortho-ro-vibrational Raman resonances, as corroborated with their frequency offset from their associated pump (average experimental shift of $12.45 \pm 0.06$ THz as compared to the theoretical value of 12.44 THz). Similarly, the green lines correspond to the Stokes and anti-Stokes sidebands of either the purple or the blue lines, and are generated by the para-ro-vibrational Raman resonance (average experimental shift of $5.30 \pm 0.20$ THz as compared to the tabulated value of 5.33 THz). Finally, the spectrum includes red lines, which, even though they can be described in terms of the Raman frequency expression $v_{l,m,n}$, are generated by more subtle mechanism than the simple cascaded Raman picture effect, as discussed below. Limiting the further discussion to only the pump laser sidebands of orders (0, $m$, $n$), we see that this part of the comb extends over 220-340 THz, and its Raman frequencies take the form of $v_{0,m,\pm1}$ with no overlapping lines from the vibrational Stokes and anti-Stokes sidebands. Remarkably, the spectrum stands out with equally spaced spectral components, with a frequency-spacing $\Delta = 1.75 \pm 0.05$ THz, which is also equal to $v_{R2} - 2v_{R3}$, the spacing of two Raman lines located at $v_{0,m+1,-1}$ and $v_{0,m,1}$. Thus, this part of the spectrum is transformed from a quasi-periodic Raman comb into a single-frequency comb whose component takes the form $\bar{v}_q = v_0 + q\Delta$, with $q$ being an integer. Figure. 1(f) shows a close-up of a portion of this spectrum, which is also representative of any range between $v_{0,m,0}$ and $v_{0,m+1,0}$. It shows that in addition to Raman blue lines ($v_{0,-2,0}$ and $v_{0,-1,0}$), and green lines ($v_{0,-2,1}$ and $v_{0,-1,-1}$), we have four red lines (labeled $v_{FWM1}$, $v_{FWM2}$, $v_{FWM3}$, $v_{FWM4}$). We start by recalling that the generation of the comb components is chronologically ordered by the Raman gain strength. Consequently, the vibrational lines (purple lines) appear first, then the ortho-ro-vibrational (blue lines) [denoted by Stage i) in Fig. 2], and finally the para-ro-vibrational Raman lines $v_{0,m,\pm1}$ (green) [Stage ii) in Fig. 2]. The latter are sidebands of the black line $v_{0,0,0}$ (pump), and blue lines $v_{0,m,0}$ (ortho-ro-vibrational Raman). This sequence was corroborated theoretically (see below). In order to determine the dominating mechanism between cascaded SRS and Kerr effect in the red lines generation, we qualitatively assessed the Raman gain in the transient regime [24-25] and Kerr gain [26]; or alternatively their associated characteristic lengths (see Appendix A: Competition between SRS and FWM). The results show that the Kerr characteristic length is 110 shorter than the Raman characteristic length, thus indicating that the generation of lines at $v_{FWM1}$, $v_{FWM2}$, $v_{FWM3}$, $v_{FWM4}$ is dominated by Kerr nonlinearity. Consequently, the red lines result from a cascaded parametric FWM process, which is further facilitated by the combination of three favorable circumstances: (i) The near-commensurability of $v_{R2}$ and $v_{R3}$, (ii) The small frequency spacing between two adjacent green lines, and (iii) The PMC weak dispersion (see below and figure 3). Below, we discuss the involved comb lines behind this wave-mixing, and the origin of such a Kerr-type nonlinearity. These red-colored lines can result from wave-mixing involving several combinations of $v_{0,m,n}$. For example, looking at the $v_{FWM2}$ spectral line 269.8 THz in Fig. 1(f), we identify several possible FWM processes to generate this line, like $v_{FWM2} = 2v_{0,-2,1} - v_{0,-1,-1}$ or $v_{FWM2} = v_{FWM1} + v_{0,-1,-1} - v_{0,-2,1}$ or $v_{FWM2} = v_{0,-2,1} + v_{0,-2,0} - v_{FWM1}$, and so on. Our aim now is to determine which of the possible FWM processes and spectral combinations lead to the generation of the red lines. Due to complicated dynamics involving high-order nonlinearities, the cascaded nature of the considered process, and the influence of phase-matching, we found it useful to start by recalling the expression for the intensity of the new frequency $v_1$ by FWM process from frequencies $v_2$, $v_3$ and $v_4$ [27]:

$$I_1 = 9\pi^2 \chi^{(3)\,2}(v_1, v_2, v_3, v_4) I_2 I_3 I_4\, v_1^2\, \varepsilon_0^2 sin^2(\Delta k z) \quad (1)$$

Where $\chi^{(3)}(v_1, v_2, v_3, v_4)$ is the frequency-dependent nonlinear susceptibility, $I_1$ to $I_4$ are the intensities at the participating frequencies, $z$ is the propagation length, and $\Delta k$ is the associated wavenumber mismatch. Expression (1) shows three possible dominating effects in the generation of red spectral lines: i) the strength of frequency-dependent nonlinear susceptibility (see Appendix B: Susceptibility in Raman-active medium); ii) the powers of the participating lines; and iii) the mismatch $\Delta k$ between the participating frequencies. Before trying to identify the leading effects in the observed FWM, we can easily disregard the nonresonant Kerr susceptibility, as the estimated characteristic length $L_{NL} = c/(2\pi. v_0. I. n_2)$ necessary for the generation of FWM sidebands by nonresonant Kerr process is $\sim 150$ m for $n_2 = 0.3 \times 10^{-8}$ cm$^2$/TW, which is much larger than the fiber length. On the other hand, we can calculate a rough estimate of Raman-induced effective value of $n_{2,Raman}$ by comparing the characteristic length observed experimentally for the red lines with that of the Kerr effect. We obtain an effective value of $n_{2,Raman} = 3 \times 10^{-8}$ cm$^2$/TW, which corresponds to the value of $5.5 \times 10^{-8}$ measured for $D_2$ [28]. In the Appendix B, we have derived a rigorous expression for the nonlinear polarization in a Raman medium using the standard adiabatic elimination of the upper Raman states. The resultant expression is cumbersome to investigate, however, for the aim of current analysis it suffices to extract the terms that are not in resonance with the Raman transition, and keep the dominant from the remaining terms. The expression of a representative term is:

$$P_{R-FWM}(t) =$$
$$\frac{N\mu_{ja}^2}{(8\pi\hbar)^3}|\mu_{ja}|^2 \sum_{q,q' \neq q} \sum_{q''} \frac{E_q^* E_{q'} E_{q''} e^{i2\pi(v_q - v_{q'} - v_{q''})t}}{(v_q - v_{q'})(v_j - v_{q'})(v_j - v_{q''})} \quad (2)$$

This Raman induced non-resonant polarization clearly shows that the polarization responsible for the FWM process increases for fields with close frequencies due to the $v_q - v_{q'}$ term appearing in the denominator (see Appendix B for notations). Based on the above, we conclude that the FWM process contributing to the red line generation stems from Raman induced enhanced nonlinearity. Second, by observing that the ratio of the values of the susceptibility $\chi^{(3)}(v_1, v_2, v_3, v_4)$ for two different possible FWM processes is determined by the ratio of the smallest difference between the participating frequencies, we can deduce that among all the possible spectral lines fulfilling the required energy conservation, only those with small frequency difference dominate. For example, we found this ratio to have a maximum value of 2 for the relevant FWM processes from neighboring lines, while mechanisms involving non-neighboring lines will be suppressed by the factor in the denominator. Further shortlisting of the dominant FWM frequency combinations can be done using the phase-matching condition. Figure. 3 shows the coherence length, $L_{coh} \equiv \pi \Delta k^{-1}$, for three representative FWM frequency combinations. The first two correspond to $v_{FWM2} = 2v_{0,-2,1} - v_{0,-1,-1}$ and $v_{FWM1} = v_{0,-2,0} + v_{0,-1,-1} - v_{0,-2,1}$, respectively, and include frequencies from only neighboring lines. The third is of type $v_{FWM1} = v_{0,-1,-1} + v_{0,0,1} - v_{0,-1,0}$, $v_{FWM1} = v_{0,-2,0} + v_{0,0,-1} - v_{0,-1,1}$ containing larger spaced frequencies.

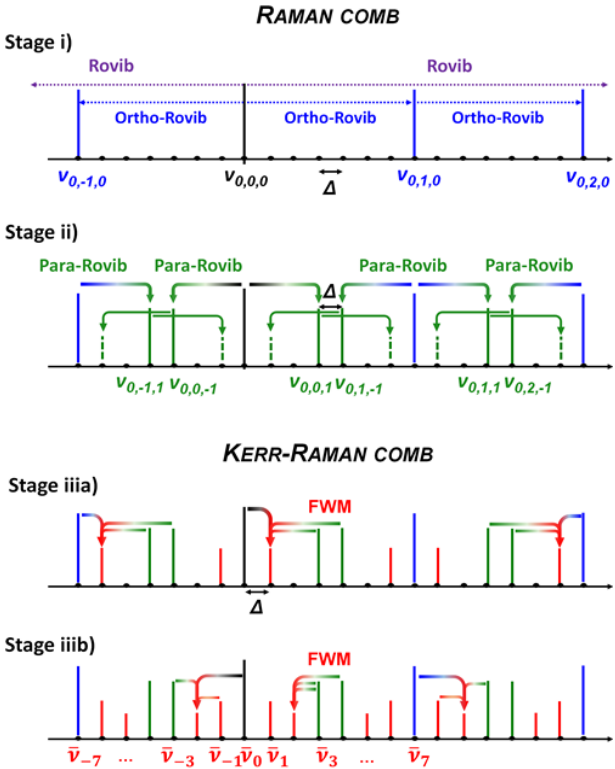

**Fig. 2**. Schematic sequence of the processes during the comb generation. The pump laser line is identified by black color, the vibrational Stokes and anti-Stokes by purple (not visible), the Stokes and anti-Stokes lines due to ortho-ro-vibrational Raman by blue, the Stokes and anti-Stokes lines due to para-ro-vibrational Raman by green, and four-wave-mixing generated sidebands by red.

The results show that for the majority of relevant processes from neighboring lines $L_{coh}$ is longer than the fiber length, so that phase mismatch is fulfilled, and does not play a significant role in selecting the FWM mechanism except for suppressing contributions from non-neighboring lines. On the other hand, the difference between the intensities of the different line series is typically one order of magnitude. Therefore, we can conclude that the power of the participating frequencies will play the dominant role in selecting the FWM process which provides the maximum contribution to the generation of new lines. Based on that, we expect that the strongest process which contributes to the generation of the $v_{FWM1}$ line is $v_{FWM1} = v_{0,-2,0} + v_{0,-1,-1} - v_{0,-2,1}$ process, while the subsequent $v_{FWM2} = v_{0,-2,1} + v_{0,-2,0} - v_{FWM1}$ and $v_{FWM2} = 2v_{0,-2,1} - v_{0,-1,-1}$ processes are mainly responsible for the generation of $v_{FWM2}$ lines. These processes are indicated as Stage iiia) and Stage iiib) in Fig. 2. Other processes certainly provide additional and coherent contribution to the abovementioned ones, since the offsets of all frequencies are roughly multiples of $\Delta$. For deuterium gas we get a clear comb-like spectrum with a step of approximately $\Delta = 1.75$ THz. The generation lines at $v_{FWM3}$ and $v_{FWM4}$ follow similar process as for $v_{FWM1}$ and $v_{FWM2}$. Finally, the spectrum shows very weak orange lines in Fig. 1(e-f), which are equally spaced from their adjacent lines by $\sim \Delta/2$. Because of the same arguments above, these lines result from further mixing of the lines. Due to their weakness, we do not find it worthwhile to identify their originating lines.

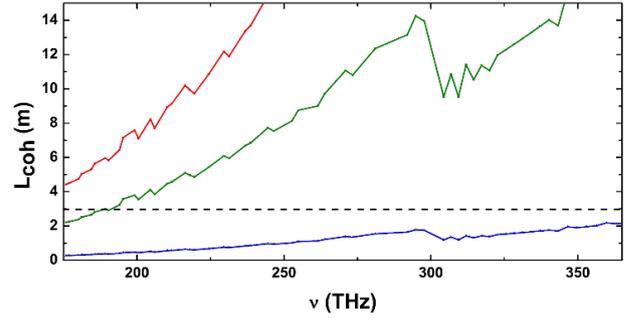

**Fig. 3**. The coherence length for the $v_{FWM2} = 2v_{0,-2,1} - v_{0,-1,-1}$ type of FWM process (red curve), for the $v_{FWM1} = v_{0,-2,0} + v_{0,-1,-1} - v_{0,-2,+1}$ type of FWM process (green curve), and for the $v_{FWM1} = v_{0,-2,0} + v_{0,0,-1} - v_{0,-1,1}$ type of FWM process (blue curve). The dashed black line indicates the 3 m fiber length.

## III. NUMERICAL AND THEORETICAL APPROACH

To corroborate the above conclusions and to achieve a deeper understanding of the processes which govern the generation of FWM sidebands in Raman-gas medium, we have developed a numerical model. The simulation of the Raman comb generation was performed by numerically solving the generalized nonlinear Schrödinger equation in the slowly varying envelope approximation which is written as [23, 29]:

$$\frac{\partial A(\omega)}{\partial z} = i \left[ k(\omega) - k(\omega_0) - (\omega - \omega_0) \frac{\partial k(\omega)}{\partial \omega}\bigg|_{\omega_0} \right] A(\omega) + \frac{i\omega}{2c\varepsilon_0} \sum_{j=2,3} P_{R,j}(\omega) \qquad (3)$$

where $A(\omega)$ is the field envelope in the spectral presentation, $k(\omega)$ is the wavenumber which includes both gas dispersion (calculated using the Sellmeyer-type expressions), the waveguide dispersion (determined from [23]), the phase mismatch, and the fiber loss spectrum. The quantity $\omega_0$ is the pump central angular frequency, and $P_{Rj}$ is the Raman polarization induced by the Raman transition $Rj$. In our case, we consider only the two ro-vibrational resonances of ortho- and para-deuterium, and ignore the vibrational transition. This choice is motivated by the large value of the vibrational Raman frequency, the higher fiber loss at $v_{\pm 1,0,0}$, and correspondingly the lower number of sidebands generated by the vibrational Stokes and anti-Stokes. In this simulation, we have deliberately neglected the nonresonant Kerr response since it is weak in comparison to nonresonant Raman response, and to determine the role of Raman in the non-resonant polarization. The Raman

polarization is given by:

$$P_{R,j}(t) = A(t)\varepsilon_0 \tilde{p}_j(\alpha_{\rho,j}\rho_{12,j} + \alpha_{w,j}w_j) \quad (4)$$

where $\tilde{p}_j$ are pressures, $\rho_{12,j}$ and $w_j$ are the off-diagonal element of the density matrix and the inversion, correspondingly, and $\alpha_{pj}$ and $\alpha_{wj}$ are coefficients which are related to the dipole moment of the Deuterium molecules. The temporal evolution of the quantities $\rho_{12,j}$ and $w_j$ is described by standard two-photon two-level Bloch equations:

$$\frac{\partial \rho_{12,j}}{\partial t} = \rho_{12,j}\left[-\frac{1}{\tau_j} + i\omega_{Rj} + \tilde{\alpha}_{1,j}|A(t)|^2\right] + \alpha_{2,j}w_j|A(t)|^2 \quad (5)$$

$$\frac{\partial w_j}{\partial t} = -\frac{w_j+1}{T_j} + \alpha_{3,j}(\rho_{12,j} - \rho^*_{12,j})|A(t)|^2 \quad (6)$$

Here $t_j$ and $T_j$ are the polarization and inversion decay times, $\alpha_{1,j}$, $\alpha_{2,j}$, and $\alpha_{3,j}$ are coefficients which describe the coupling between electric field and the state of a molecule. All of the above coefficients are determined phenomenologically, as described in *e.g.* [29].

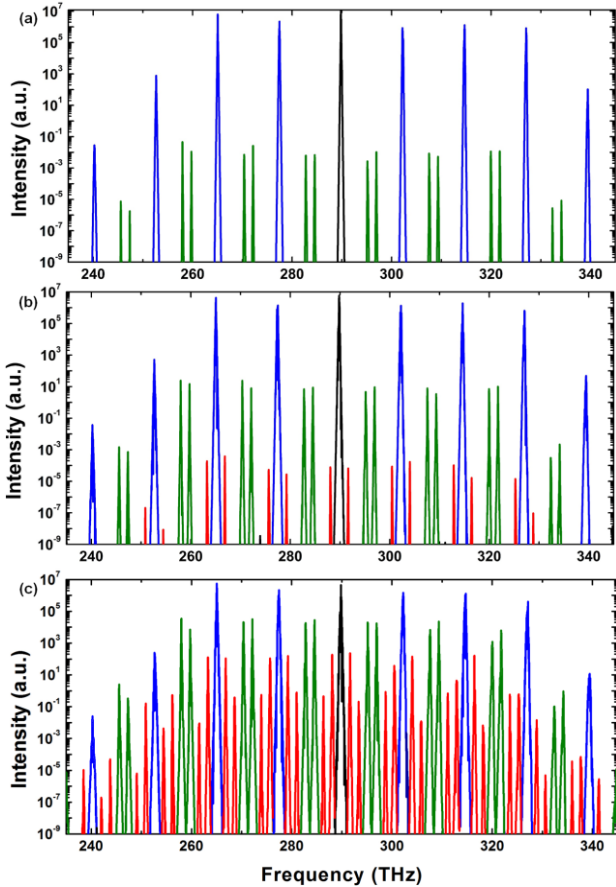

**Fig. 4**. (a-c) Evolution of the numerical output spectra for the case of a coupled pump power of 12.8 W (a), 14.4 W (b) and 16 W (c). The fiber length was 3 m long, the $D_2$ gas pressure was 40 bar. The pump laser line is shown with black color, the ortho-ro-vibrational shifts with blue color, the para- ro-vibrational with green color, and all the FWM-generated sidebands with red color.

A single carrier frequency along with envelope varying on a few-fs time scale were used to treat all the comb components simultaneously and self-consistently. Figure 4 shows the calculated evolution of the output spectra for the case of coupled pump power of 12.8 W [Fig. 4(a)], 14.4 W [Fig. 4(b)] and 16 W [Fig. 4(c)] respectively. These numerical output spectra present the same pattern as the experimental one and in a good agreement with the generation sequence of Fig. 2. The color scheme is the same as in Fig. 1, with the pump laser line $v_{0,0,0}$ shown by black color, the ortho-$D_2$ $v_{0,m,0}$ lines by blue, the para-$D_2$ $v_{0,m,\pm1}$ lines by green, and all the FWM sidebands $v_{FWM}$ shown by red color. Besides the very good agreement between the numerical and experimental data, the main conclusion of the numerical approach is the mechanism and sequence of processes, which yield the comb in a molecular Raman gas medium. In the simulation, as the propagation distance increases, ortho-$D_2$ ro-vibrational comb (blue) appears first during Stage i). The lines of this comb are sufficiently strong, that during the Stage ii) para-$D_2$ ro-vibrational comb (green) appears on both sides of each line in ortho-$D_2$ ro-vibrational comb. Figure. 4(a) shows perfectly the generated and classic Raman comb only based on the pure Raman transitions. After that, as a third step, FWM processes set in. They include the $v_{FWM1} = v_{0,m,0} + v_{0,m+1,-1} - v_{0,m,1}$ type of FWM [Stage iiia) and Fig. 4(b)] as explained above, followed by $v_{FWM2} = v_{0,m,1} + v_{0,m,0} - v_{FWM1}$ and $v_{FWM2} = v_{0,m,1} - v_{0,m+1,-1}$ [Stage iiib) and Fig. 4(c)]. This sequence is a cascaded one, since the $v_{FWM1}$ is required for the $v_{FWM2}$ to be generated. Figure 4 also confirms such a sequence, where one can see that the $v_{FWM1}$ lines are stronger than $v_{FWM2}$. Finally, very weak orange lines are not visible in the numerical simulations. This sequence of the elementary processes, in particular the mechanism of the generation of the red FWM lines, agrees with the analysis presented in the previous section, which shows that the small and single frequency spacing is related to the two-step Raman process, followed by cascaded parametric FWM, which is based on Kerr-type nonlinearity provided by the medium Raman response.

## IV. CONCLUSION

In conclusion, we reported on the generation of a high power, broadband and single-frequency spaced comb based on a sequence of SRS and parametric four-wave in a multiple Raman resonance molecular gas medium. The comb frequency spacing $\Delta$ results from the fact that two Raman frequencies of the gas are multiple of $\Delta$. This new optical comb generation process unlocks several limitations of current comb generation techniques. It can combine the multi-octave bandwidth of HC-PCF based Raman combs with single and smaller frequency spacing than in current spectra used for waveform synthesis [1] and attosecond pulse generation [9].

Also, unlike with micro-resonator or SRS comb, the frequency spacing can be actively controlled. This can be achieved for example by exciting a Raman medium, having at least two nearly commensurable Raman resonance frequencies, with two pump lasers whose frequency difference can be tuned to a value equal to a fraction of one of the two commensurable resonance frequencies. Finally, the results represent an important milestone towards exploring this Raman-Kerr comb into a frequency ruler [8]. However, unlike the presented pumping scheme, this requires specific pumping schemes such as the use of pump laser with stabilized carrier-envelop phase, and with sufficiently high repetition rate so the required small number of Raman excited molecules survive the duration between two adjacent pump pulses [30].

## ACKNOWLEDGMENTS

Funding: Agence Nationale de la Recherche (ANR) (PhotoSynth, Labex SigmaLim, UV factor); Région Limousin; Air Force Office of Scientific Research (AFOSR) (FA9+550-14-1-0024); National Science Foundation (NSF) (PHY-1068865). The authors thank the PLATINOM platform in their help in the fiber fabrication.

## APPENDIX A: COMPETITION BETWEEN SRS AND FWM

In order to have a qualitative assessment on the dominant process between SRS and FWM in the generation of the red line (Fig. 1), we proceed by estimating their effective gain or alternatively their generation characteristic lengths. For both SRS and Kerr effect, we can write the following for Raman net gain:

$$G_R = g_R(P_R/A)L = LL_R^{-1} \quad (A1)$$

with

$$g_R(P_R/A) = L_R^{-1} \quad (A2)$$

and for the Kerr process, the characteristic length required to reach the power $P_{si}$ can be derived as [27]:

$$L_K = \frac{\lambda}{2\pi n_2}\sqrt{\frac{A^2 P_{si}}{P_K^3}} \quad (A3)$$

Although the generation of a new frequency from three existing components is not a gain-type process, the following gain coefficient can be formally associated with the relevant four-wave mechanism:

$$g_K = L_K^{-1} \quad (A4)$$

Here, $P_R$ and $P_K$ are the optical powers of the exciting fields involved in SRS and FWM respectively. In the expression of the Kerr gain, we ignore the phase matching condition between the fields in the FWM. $L_R$ and $L_K$ are the generation characteristic lengths for the process of SRS and Kerr respectively. The stronger the generation process the shorter the characteristic length. Consequently, finding out whether FWM is dominant or not is determined by the ratio ($L_K/L_R$). The quantity $\lambda$ is the wavelength of the pumping field. $n_2$ is the nonlinear index of the Raman driven $D_2$, and $g_R$ is the Raman gain coefficient.

Recalling that in the transient regime, the Raman gain coefficient $g_R$ is equal to $g_{R,t}$ which is related to the steady-state Raman gain coefficient $g_{R,ss}$ $g_{R,ss}$ by [24]:

$$g_{R,t} = R \times g_{R,ss} \quad (A5)$$

with

$$R = 2\sqrt{\frac{\Gamma \tau}{g_{R,ss}(P_R/A)L}} \quad (A6)$$

And, the pulse duration of the Stokes, $\tau_S$ is given by [25]:

$$\tau_S = \sqrt{\frac{\tau}{16 g_{R,ss} \Gamma (P_R/A) L}} \quad (A7)$$

Here, $\Gamma$ is the Raman dephasing rate, $\tau$ is the pulse duration of the Raman pumping field.

Let us take the case of cascade SRS generation of one of the spectral lines of interest, the line at frequency $v_{0,-1,-2}$ (labelled FWM1). The latter is the second order Stokes of the blue line at $v_{0,-1,0}$ or the first order Stokes of the green line at $v_{0,-1,-1}$. Using the above expressions of the gain coefficient and the pulse duration (equations (A6) and (A7)), we readily find the Raman gain coefficient for the generation of the red line $v_{0,-1,-2}$. Setting the average power of the blue line $v_{0,-1,0}$ to 1.6 W, and that of the green line $v_{0,-1,-1}$ to 0.16 W, which are consistent with the measured output spectrum when 16 W of the laser power is sent to the fiber, we find $g_{R,t} \approx 10^{-6}$ cm.GW$^{-1}$ and $L_R \approx 495$ cm for the generation of the red line $v_{0,-1,-2}$.

Similarly, we estimate the Kerr gain and characteristic length by setting the pumping average power to 1.6 W, and taking $n_2 = 3 \times 10^{-8}$ cm$^2$/TW, the value estimated as explained in the text of the paper [28]. We found $g_K \approx 3.8 \times 10^{-6}$ cm.GW$^{-1}$ and $L_K \approx 4.5$ cm.

These approximate calculations show that $L_R$ is 110 times larger than $L_K$, thus indicating that Kerr dominates the red line $v_{0,-1,-2}$. Similar approach shows that this is valid for all red lines.

## APPENDIX B: SUSCEPTIBILITY IN RAMAN-ACTIVE MEDIUM

The full expression for the susceptibility in the Raman-active medium is discussed. Our derivation of the expression for the nonlinear susceptibility of the Raman medium follows the general line and denotations of Ref. [29]. We begin the derivation of this expression by considering a system which can be described by a three-level system with two lower levels $|a\rangle$ and $|b\rangle$ with energies $\hbar\omega_a$ and $\hbar\omega_b$, $\omega_a < \omega_b$, and higher level $|j\rangle$ with energy $\hbar\omega_j$. Without the loss of generality we $\omega_a = 0$ and $\omega_a = \omega_R$, where $\omega_R$ is the Raman shift. The higher level $j$ can be understood as a combination of multiple higher levels. The external field has three frequency components:

$$E(t) = \frac{1}{2}\sum_{j=1,3} E_j e^{-i\omega_j t} + c.c. \quad (B1)$$

which couple the energy levels of the system through the Hamiltonian

$$H = \begin{pmatrix} 0 & 0 & V_{aj} \\ 0 & \hbar\omega_R & V_{bj} \\ V_{ja} & V_{jb} & \hbar\omega_j \end{pmatrix} \quad (B2)$$

Here

$$V_{aj} = V_{ja}^* = -\mu_{aj}E \quad (B3)$$
$$V_{bj} = V_{jb}^* = -\mu_{bj}E \quad (B4)$$

As one can see from these expressions, the levels $|a\rangle$ and $|j\rangle$, as well as $|a\rangle$ and $|j\rangle$ are coupled by electric dipole transition, while the transition between $|a\rangle$ and $|b\rangle$ is forbidden.

In this case, the amplitudes $C_n$ of the wavefunction of the system

$$|\psi\rangle = \sum_{n=a,b,j} C_n e^{-i\omega_n t} |n\rangle \qquad (B5)$$

satisfy the following equations which follow from the Schrödinger equation:

$$\frac{\partial C_n}{\partial t} = -\frac{i}{\hbar}\sum_k e^{-(\omega_n - \omega_k)t} V_{nk} C_k \qquad (B6)$$

This set of linear equations is solved by usual methods, utilizing the adiabatic elimination of $C_j$ [29]. The details of the derivation are standard and too cumbersome to be presented here. After the expression for the coefficients $C_n$ are obtained, the polarization of the medium can be expressed through

$$P = N\sum_{n=a,b} \mu_{ja} C_a C_j^* e^{-i(\omega_a - \omega_j)} + c.c. \qquad (B7)$$

$N$ being the particle density.
The final expression for the polarization reads

$$P(t) = \frac{N\mu_{ja}^2}{64\hbar^3}\left(|\mu_{ja}|^2 A + |\mu_{jb}|^2 B\right) \qquad (B8)$$

Where

$$A = \left\{\sum_{q,q'}\left[-\frac{E_q E_{q'} e^{-i(\omega_q - \omega_{q'})t}}{(\omega_q + \omega_{q'})(\omega_j - \omega_{q'})} + \frac{E_q^* E_{q'}^* e^{-i(\omega_q - \omega_{q'})t}}{(\omega_q - \omega_{q'})(\omega_j - \omega_{q'})}\right] + \right.$$
$$\sum_{q,q'\neq q}\left[-\frac{E_q E_{q'}^* e^{-i(\omega_q - \omega_{q'})t}}{(\omega_q - \omega_{q'})(\omega_j + \omega_{q'})} + \frac{E_q^* E_{q'} e^{-i(\omega_q - \omega_{q'})t}}{(\omega_q - \omega_{q'})(\omega_j - \omega_{q'})}\right] + $$
$$\left. c.c.\right\} \times \left\{\sum_{q''}\left[\frac{E_{q''} e^{-i\omega_{q''}t}}{\omega_j - \omega_{q''}} + \frac{E_{q''}^* e^{i\omega_{q''}t}}{\omega_j + \omega_{q''}}\right] + c.c.\right\} \qquad (B9)$$

And

$$B = \left\{\sum_{q,q'}\left[-\frac{E_q E_{q'} e^{-i(\omega_q + \omega_{q'})t}}{(\omega_R - \omega_q - \omega_{q'})(\omega_j - \omega_{q'})} + \right.\right.$$
$$\left.\frac{E_q^* E_{q'}^* e^{i(\omega_q + \omega_{q'})t}}{(\omega_R + \omega_q + \omega_{q'})(\omega_j + \omega_{q'})}\right] +$$
$$\sum_{q,q'\neq q}\left[-\frac{E_q E_{q'}^* e^{-i(\omega_q - \omega_{q'})t}}{(\omega_R - \omega_q + \omega_{q'})(\omega_j + \omega_{q'})} + \right.$$
$$\left.\frac{E_q^* E_{q'} e^{i(\omega_q - \omega_{q'})t}}{(\omega_R + \omega_q - \omega_{q'})(\omega_j - \omega_{q'})}\right] + c.c.\right\} \times \left\{\sum_{q''}\left[\frac{E_{q''} e^{-i\omega_{q''}t}}{\omega_j - \omega_R - \omega_{q''}} + \right.\right.$$
$$\left.\frac{E_{q''}^* e^{i\omega_{q''}t}}{\omega_j - \omega_R + \omega_{q''}} + \frac{E_{q''}^* e^{i\omega_{q''}t}}{\omega_j - \omega_{q''}} + \frac{E_{q''} e^{-i\omega_{q''}t}}{\omega_j + \omega_{q''}}\right] + c.c.\right\} \qquad (B10)$$

In this expression, indices $q, q', q''$ denote pump fields and can take values from 1 to 3. In deriving this expression, we have neglected the decay rates which are for our case much lower than the few-THz difference between frequencies. This expression includes all contributions provided by the Raman nonlinearity, including Raman contributions to self- and cross-phase modulation and in particular four-wave mixing. Equation (B10) contains in total 1440 terms, therefore in the main text we present a susceptibility corresponding to only one of these contributions. Since the values of dipole momenta $\mu_{ja}$ and $\mu_{ja}$ are similar, the dominant terms are determined by the smallest value of the denominator. For typical conditions, $\omega_j \approx \omega_R$ and $\omega_j \approx \omega_q$, therefore the factors in the denominator which contain $\omega_j$ will have similar value for all terms, with terms containing $\omega_j - \omega_q$ in the denominator being larger. On the other hand, the terms in A which contain in the denominator the combination $\omega_q - \omega_{q'}$, and the terms in B which contain in the denominator combinations $\omega_R \pm (\omega_q - \omega_{q'})$, are larger than other terms and will provide dominant contribution to the respective nonlinear processes. Considering the above arguments, we identify the dominant contributions $A_{dom}$ and $B_{dom}$ to A and B as follows:

$$A_{dom} = \left\{\sum_{q,q'\neq q} \frac{E_q^* E_{q'} e^{i(\omega_q - \omega_{q'})t}}{(\omega_q - \omega_{q'})(\omega_j - \omega_{q'})} + \right.$$
$$\left. c.c.\right\}\left\{\sum_{q''}\frac{E_{q''} e^{-i\omega_{q''}t}}{\omega_j - \omega_{q''}} + c.c.\right\} \qquad (B11)$$

And

$$B_{dom} = \left\{\sum_{q,q'} \frac{E_q E_{q'} e^{i(\omega_q - \omega_{q'})t}}{(\omega_R - \omega_q - \omega_{q'})(\omega_j - \omega_{q'})} + \right.$$
$$\sum_{q,q'} \frac{E_q^* E_{q'} e^{i(\omega_q - \omega_{q'})t}}{(\omega_R + \omega_q - \omega_{q'})(\omega_j - \omega_{q'})} + c.c.\right\} \times \left\{\sum_{q''}\left[\frac{E_{q''} e^{-i\omega_{q''}t}}{\omega_j - \omega_R - \omega_{q''}} + \right.\right.$$
$$\left.\left.\frac{E_{q''}^* e^{i\omega_{q''}t}}{\omega_j - \omega_{q''}}\right] + c.c.\right\} \qquad (B12)$$